\documentclass{myaa}

\usepackage[varg]{txfonts}

\usepackage{natbib}
\usepackage{graphicx}
\usepackage{epstopdf}
\usepackage{amssymb}
\usepackage{float}

\begin{document}

\title{Theoretical model of the outer disk of TW Hya presently forming in-situ planets and comparison with models of AS 209 and HL Tau}

\author{Dimitris M. Christodoulou\inst{1,2}  
\and 
Demosthenes Kazanas\inst{3}
}

\institute{
Lowell Center for Space Science and Technology, University of Massachusetts Lowell, Lowell, MA, 01854, USA.\\
\and
Dept. of Mathematical Sciences, Univ. of Massachusetts Lowell, 
Lowell, MA, 01854, USA. \\ E-mail: dimitris\_christodoulou@uml.edu\\
\and
NASA/GSFC, Laboratory for High-Energy Astrophysics, Code 663, Greenbelt, MD 20771, USA. \\ E-mail: demos.kazanas@nasa.gov \\
}


\def\gsim{\mathrel{\raise.5ex\hbox{$>$}\mkern-14mu
                \lower0.6ex\hbox{$\sim$}}}

\def\lsim{\mathrel{\raise.3ex\hbox{$<$}\mkern-14mu
               \lower0.6ex\hbox{$\sim$}}}

\abstract{
We fit an isothermal oscillatory density model to the outer disk of TW Hya in which planets have presumably already formed and they are orbiting within four observed dark gaps. At first sight, this 52 AU small disk does not appear to be similar to our solar nebula; it shows several physical properties comparable to those in HL Tau (size $R_{\rm max}=102$ AU) and very few similarities to AS 209 ($R_{\rm max}=144$ AU). We find a power-law density profile with index $k=-0.2$ (radial densities $\rho(R) \propto R^{-1.2}$) and centrifugal support against self-gravity so small that it virtually guarantees dynamical stability for millions of years of evolution to come. Compared to HL Tau, the scale length $R_0$ and the core size $R_1$ of TW Hya are smaller only by factors of $\sim$2, reflecting the disk's half size. On the opposite end, the Jeans frequency $\Omega_J$ and the angular velocity $\Omega_0$ of the smaller core of TW Hya are larger only by factors of $\sim$2. The only striking difference is that the central density ($\rho_0$) of TW Hya is 5.7 times larger than that of HL Tau, which is understood because the core of TW Hya is only half the size ($R_1$) of HL Tau and about twice as heavy ($\Omega_J$). In the end, we compare the protostellar disks that we have modeled so far.}
\keywords{planets and satellites: dynamical evolution and stability---planets and satellites: formation---protoplanetary disks}

\authorrunning{ }
\titlerunning{Planet formation in the outer disk of TW Hya}

\maketitle

\section{Introduction}\label{intro}

In previous work \citep{chr19a}, we presented isothermal models of the solar nebula capable of forming protoplanets long before the protosun is actually formed by accretion processes. This entirely new ``bottom-up'' formation scenario is currently observed in real time by the latest high-resolution ($\sim$1-5~AU) observations of many protostellar disks by the ALMA telescope \citep{alm15,and16,rua17,lee17,lee18,mac18,ave18,cla18,kep18,guz18,ise18,zha18,dul18,fav18,har18,hua18,per18,kud18,lon18,pin18,vdm19}.   

The ALMA/DSHARP observations show many circular protostellar disks with circular dark gaps presumably carved out by protoplanets that have already been formed at a time long long before accretion/dispersal processes will dissipate their disks. Very few disks show asymmetries and spiral arms, signs of setting-in instabilities \citep{per18,hua18,vdm19}. Motivated by these observations, we have produced models of the disks of AS 209 (seven gaps), HL Tau (seven gaps), and RU Lup (4 gaps) \citep{chr19b,chr19c}. In this work, we apply the same theoretical model to the observed disk of TW Hya, a prototypical young system observed by ALMA/DSHARP \citep{alm15,hua18,hua18b}. 

TW Hya is unusual in that it shows a inner dark gap at 1 AU, and the next gap, D26, lies at 25.62 AU. It is obvious that the inner disk is not resolved, but then the outer disk shows clearly four dark gaps in a configuration that is not easy to model: gap D42 is 10 AU away from D32, but the next outer gap, D48, lies 4 AU closer to D42. The disk of TW Hya is relatively small in extent ($R_{\rm max}=52$ AU). It seems that the observations managed to find only the dark gaps in the outer disk (beyond 25 AU). We think that there may be more planets between 1AU and 25 AU, but we have no way of modeling them presently. This also means that the core region of TW Hya will be large, making room for the yet undetected planets. So we ignore the 1 AU gap, and we focus our models to the outer disk with its four pronounced dark gaps.

The analytic (intrinsic) and numerical (oscillatory) solutions of the isothermal Lane-Emden equation \citep{lan69,emd07} with differential rotation, and the resulting model of the midplane of the gaseous disk have been described in detail in \cite{chr19a} for the solar nebula. Here, we apply in \S~\ref{models2} the same model to the four outer dark gaps of TW Hya, and we compare the best-fit results against AS 209 and, more importantly, HL Tau \citep{chr19b}. In \S~\ref{disc}, we summarize our results and we discuss the ALMA-observed disks that we have modeled so far.

\section{Physical Models of the TW Hya Protostellar Disk}\label{models2}

The numerical integrations that produce oscillatory density profiles were performed with the \textsc{Matlab} {\tt ode15s} integrator \citep{sha97,sha99} and the optimization used the Nelder-Mead simplex algorithm as implemented by \cite{lag98}. This method (\textsc{Matlab} routine {\tt fminsearch}) does not use any numerical or analytical gradients in its search procedure which makes it extremely stable numerically, albeit somewhat slow. The boundary conditions for the oscillatory density profiles are, as usual, $\tau(0)=1$ and $[d\tau/dx](0)=0$, where $\tau$ and $x$ are the dimensionless values of the density and the radius, respectively.

\begin{figure}
\begin{center}
    \leavevmode
      \includegraphics[trim=0.2 0.2cm 0.2 0.2cm, clip, angle=0, width=10 cm]{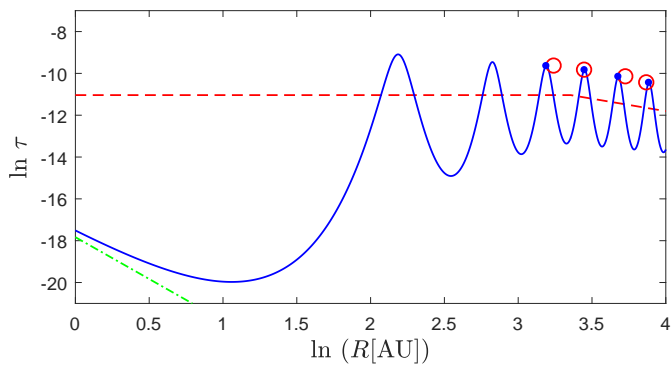}
      \caption{Equilibrium density profile for the midplane of TW Hya outer disk that has already formed at least four annular dark gaps (presumably protoplanets) \citep{hua18}. The best-fit parameters are $k=-0.2$, $\beta_0=0.00401$, and $R_1=28.67$~AU. The radial scale length of the disk is $R_0=0.004100$~AU. The Cauchy solution (solid line) has been fitted to the outer dark gaps of TW Hya (Table~\ref{table1}) so that its density maxima (dots) correspond to the observed orbits of the protoplanets (open circles). The density maximum corresponding to the location of the fourth maximum was scaled to a distance of 31.5~AU of the D32 gap. The first two density peaks (at 8.9 AU and 16.9 AU) had to be left empty. The mean relative error of the fit is 5.1\%, most of it coming from gaps D26 and D42 (Table~\ref{table1}). The intrinsic analytical solution (dashed line) and the nonrotating analytical solution (dash-dotted line) are also shown for reference. 
\label{fig1}}
  \end{center}
\end{figure}

\begin{table}
\caption{Radii of dark gaps in AS 209, HL Tau, and TW Hya \citep[from Table 1 of][]{hua18}}
\label{table1}
\begin{tabular}{ll|ll|ll}
\hline
Gap   & AS 209     & Gap    &  HL Tau & Gap  &  TW Hya \\
Name & $R~(AU)$ & Name  &  $R~(AU)$ & Name & $R~(AU)$ \\
\hline
D9   &   8.69   &   D14   &   13.9 & D1 & 1 \\
D24  &  23.84  &   D34   &   33.9 & D26 &  25.62 \\
D35  &  35.04  &   D44   &   44 & D32 & 31.5 \\
D61  &  60.8    &   D53   &   53 & D42 &  41.64\\
D90  &  89.9    &   D67   &   67.4 & D48 & 48 \\
D105 &  105.5  &   D77   &   77.4 & & \\
D137 &  137     &   D96   &   96 & & \\
\hline
\end{tabular}
\end{table}

\begin{table*}
\caption{Comparison of the protostellar disk model of TW Hya against AS 209 and HL Tau}
\label{table2}
\begin{tabular}{lllll}
\hline
Property & Property & AS 209 & HL Tau & TW Hya \\
Name     & Symbol (Unit) & Best-Fit Model & Best-Fit Model & Best-Fit Model \\
\hline
Density power-law index & $k$                                          &   $0.0$  	     & $0.0$   & $-0.2$ \\
Rotational parameter & $\beta_0$                                &    0.0165	       &  0.00562 &  0.00401 \\
Inner core radius & $R_1$ (AU)                              &   6.555  	       &  52.04  &  28.67 \\
Outer flat-density radius & $R_2$ (AU)                              &   68.96        	   &  90.55  & $\cdots$ \\
Scale length & $R_0$ (AU)    &   0.01835  &  0.009813  & 0.004100 \\
Equation of state & $c_0^2/\rho_0$ (${\rm cm}^5 {\rm ~g}^{-1} {\rm ~s}^{-2}$) & $6.32\times 10^{16}$ & $1.81\times 10^{16}$ &  $3.15\times 10^{15}$   \\
Minimum core density for $T=10$~K, $\overline{\mu} = 2.34$ & $\rho_0$ (g~cm$^{-3}$)         &    $5.62\times 10^{-9}$   			&   $1.97\times 10^{-8}$ &  $1.13\times 10^{-7}$  \\
Isothermal sound speed for $T=10$~K, $\overline{\mu} = 2.34$ & $c_0$ (m~s$^{-1}$) & 188 & 188 & 188 \\
Jeans gravitational frequency & $\Omega_J$ (rad~s$^{-1}$)    &    $4.9\times 10^{-8}$ & $9.1\times 10^{-8}$  &  $2.2\times 10^{-7}$ \\
Core angular velocity & $\Omega_0$ (rad~s$^{-1}$)    &    $8.0\times 10^{-10}$ 	& $5.1\times 10^{-10}$  & $8.7\times 10^{-10}$  \\
Core rotation period & $P_0$ (yr)                                 &    249 	   			&  390  & 228 \\
Maximum disk size & $R_{\rm max}$ (AU)                &    144 	   			&   102 & 52 \\
\hline
\end{tabular}
\end{table*} 

\subsection{Best-Fit models of TW Hya}\label{model1}

The radii of the five known dark gaps in TW Hya are shown in Table~\ref{table1}, but the 1 AU gap will be ignored in what follows. In Fig.~\ref{fig1}, we show the best optimized fit to the four outer dark gaps of TW Hya. In these models, we have used three free parameters ($k$, $\beta_0$, and $R_1$) because the disk is relatively small and it does not need a flat-density outer region. The mean relative error of the fit is 5.1\% and it comes from gaps D26 and D42 (Table~\ref{table1}).

The physical properties of the best-fit TW Hya model are listed in Table~\ref{table2} along with the best models of AS 209 and HL Tau. It is obvious that TW Hya is more similar to HL Tau. TW Hya and AS 209 share effectively only two common properties in their values of $k$ and $\Omega_0$. The agreement in $\Omega_0$ is a coincidence as the core of AS 209 is much smaller in size (by a factor of 4.4). A detailed comparison between the disks of TW Hya and HL Tau is discussed below.

\begin{figure}
\begin{center}
    \leavevmode
      \includegraphics[trim=0.2 1.5cm 0.2 1.5cm, clip, angle=0, width=10 cm]{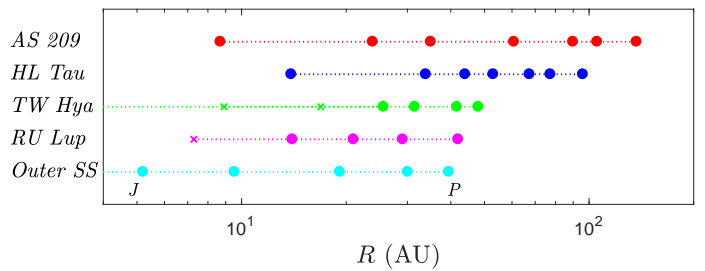}
      \caption{Schematic diagram of the ALMA-observed dark gaps (dots) that we have modeled so far. The crosses represent the empty density peaks in which no dark gaps have been observed yet. It is evident that TW Hya is about similar to RU Lup (accounting for the two empty density peaks) and that RU Lup is similar to our outer solar system (``{\it Outer SS}'') from Jupiter ({\it J}) to Pluto ({\it P}). But the physical parameters of TW Hya and RU Lup (Table~\ref{table2}) differ, in fact they indicate similar properties between TW Hya and HL Tau.
\label{fig_disks}}
  \end{center}
\end{figure}

\subsection{Comparison between the best-fit models of TW Hya and HL Tau}\label{comp}

The TW Hya power-law density profile with index $k=-0.2$ is close to the value of $k=0.0$ of the other two disks (Table~\ref{table2}). Centrifugal support against self-gravity is extremely low ($\beta_0\simeq 0.004$) for TW Hya, guaranteeing the disk's long-term dynamical stability \citep{chr95}. The TW Hya model compares well to the best-fit HL Tau model: Its scale length $R_0$ and its core size $R_1$ are smaller only by factors of 1.8-2.4. On the other hand, the Jeans frequency $\Omega_J$ and the angular velocity $\Omega_0$ of the smaller core of TW Hya are larger by factors of 1.7-2.4. These scalings do make sense.

There is, however, a striking difference in the central density $\rho_0$ of TW Hya; it is about 6 times larger than that of HL Tau, and this is understood as follows: the disk of TW Hya is only half the size of HL Tau ($R_{\rm max}=52$ AU versus 102 AU) and 2.4 times heavier (see the $\Omega_J$ values in Table~\ref{table2}).

\section{Summary}\label{disc}

In \S~\ref{models2}, we presented our best-fit isothermal differentially-rotating protostellar models of TW Hya observed by ALMA/DSHARP \citep{alm15,hua18,hua18b,guz18,zha18}. This model shows four dark gaps in the outer disk (Table~\ref{table1}), and it is widely believed that protoplanets have already formed and curved out these gaps in the observed outer disk. The best-fit model is depicted in Figure~\ref{fig1} and a comparison of its physical properties versus AS 209 and HL Tau is shown in Table~\ref{table2}. The physical properties of TW Hya are much closer to those of HL Tau than those of AS 209.

In Fig.~\ref{fig_disks}, we show a schematic diagram of dark gaps (dots) in the ALMA-observed disks that we have modeled so far and we also included our solar nebula. We have also included the empty density peaks predicted by some best-fit models and these are denoted in the figure by crosses. No dark gaps have been observed in these positions yet, so these comparisons are very much model dependent. The picture that we have formed is as follows:
\begin{enumerate}
\item The gap arrangement of HL Tau is roughly similar to that of AS 209 than any other disk. In fact, the two disks also share several physical properties (Table~\ref{table2}), except for the inner core radius $R_1$.
\item The TW Hya gaps, including the two empty density peaks, appear to have a similar arrangement to the gaps of RU Lup (Fig.~\ref{fig_disks}), but in physical properties, only the cores ($R_1$ and $\Omega_0$) of the disks are similar. On the other hand, several of the physical properties of TW Hya are much more similar to those of HL Tau (Table~\ref{table2}).
\item RU Lup definitely looks like our outer solar system, in particular, the region between Jupiter and Pluto \citep{chr19c}. Both planetary arrangements also share about the same size ($R_{\rm max}\simeq 50$ AU) and most other physical properties. Only the cores ($R_1$ and $\Omega_0$) of the disks are different, but this is because the core of RU Lup is not adequately resolved by ALMA (whereas the core of our solar nebula is populated with the four terrestrial planets).
\end{enumerate}
From this comparison, we have two ALMA-observed disks in which the dark gap arrangements in their outer regions appear to be similar to the outer planets in our solar system: RU Lup and TW Hya (but the physical properties of RU Lup are much closer to those of the solar system). This is a good start in our efforts to identify extremely young exoplanetary systems similar to our own much older solar system.

\end{document}